\documentstyle[preprint,aps]{revtex}

\begin{document}
\title{The non-retarded dispersive force between an electrically polarizable atom
and a magnetically polarizable one}
\author{C. Farina\thanks{%
farina@if.ufrj.br}, F.C. Santos\thanks{%
filadelf@if.ufrj.br} and A.C. Tort\thanks{%
tort@if.ufrj.br}}
\address{Instituto de F\'{i}sica Universidade Federal do Rio de Janeiro, Caixa Postal%
\\
68528, 21945-970 Rio de Janeiro RJ, Brazil}
\date{\today}
\maketitle
\pacs{23.23.+x, 56.65.Dy}

\begin{abstract}
Using perturbative QED we show that, while the retarded dispersive force
between an electrically polarizable atom and a magnetically polarizable one
is proportional to $1/r^{8}$, where $r$ is the distance between the atoms,
the non-retarded force is proportiaonal to $1/r^{5}$. This is a rather
surprising result that should be contrasted with the dispersive van der
Waals force between two electrically polarizable atoms, where the retarded
force is also proportional to $1/r^{8}$, but the non-retarded force is
proportional to $1/r^{7}$.
\end{abstract}

Non-retarded dispersive forces between two electrically polarizable atoms
were calculated in detail for the first time by London \cite{London30} who
applied perturbation theory in ordinary quantum mechanics and showed that
this force is proportional to $1/r^{7}$, where $r$ is the distance between
the two atoms. However, when $r$ is large compared with the transition
wavelengths involved, retardation effects in the propagation of the
electromagnetic interaction must be taken into account. Casimir and Polder 
\cite{CasPol48} investigated the influence of retardation on the London-van
der Waals forces making use of perturbative QED. They were motivated by a
conjecture made by Verwey and Overbeek \cite{VerOver48} who suggested that
the interatomic force should fall faster than $1/r^{7}$ at large distances,
otherwise experimental data for some colloidal systems and the theoretical
predictions would not agree. Casimir and Polder found that the retarded
dispersive force between two atoms is proportional to $1/r^{8}$. Therefore,
the influence of retardation is to change the exponent in the power law of
the force by one unit. This kind of change also occurs when we deal with the
force between an electrically polarizable atom and a perfectly conducting
wall. While the non-retarded force, which is valid for short distances from
the wall, is proportional to $1/r^{4}$ (basically the force between the
induced atomic dipole and its image), the retarded force is proportional to $%
1/r^{5}$ \cite{CasPol48,Milloni}.

In this letter, we willl investigate the non-retarded force between two
non-similar atoms: one of them electrically polarizable and the other one
magnetically polarizable.We will show that a quite unexpected result is
obtained, namely: while the retarded potential (force) between them is
proportional to $1/r^{7}\ $($1/r^{8})$, the non-retarded potential (force)
is proportional to $1/r^{4}$ ($1/r^{5}$). This situation must be compared
with the $1/r^{6}\;$($1/r^{7}$) power law for the non-retarded potential
(force) between two electrically polarizable atoms. As far as the authors
are aware of this striking feature has never been investigated before.

The retarded interaction energy $U(r)$ between two atoms endowed with both
electrical and magnetical polarization was discussed in detail by Feinberg
and Sucher \cite{Feinberg68}, and also Boyer \cite{Boyer69}. The result is 
\begin{equation}  \label{UBoyer}
U(r)=\left[ -23\left( \alpha _{1}\alpha _{2}+\beta _{1}\beta _{2}\right)
+7\left( \alpha _{1}\beta _{2}+\alpha _{2}\beta _{1}\right) \right] \frac{%
\hbar c}{4\pi r^{7}}\,,  \label{boyer}
\end{equation}
where in Eq. (\ref{boyer} ) $\alpha _{i}$ and $\beta _{i}$ ($i=1,2$) are,
respectively, the (static) electric polarizability and the (static) magnetic
polarizability of atom $i$. If in Eq. (\ref{boyer}) we set $\beta _{1}=\beta
_{2}=0$, we will recover Casimir and Polder's result \cite{CasPol48}.

In order to show that the non-retarded dispersion van der Waals interaction
potential for the case at hand obeys a $1/r^{4}$ law and not the $1/r^{6}$
law as one would naively think we shall use perturbative QED. We will follow
a procedure analogous to that found in \cite{Milloni}.

Consider two atoms, $A$ and $B$ and assume that atom $A$ is electrically
polarizable, while atom $B$ is magnetically polarizable. In an obvious
notation, the change in the energy level of atom $A$ is given by: 
\begin{equation}
\left\langle 0\left| \hat{W}\right| 0\right\rangle =-{\frac{1}{2}}%
\sum_{\sigma }\alpha _{A}(\omega _{\sigma })\left\langle 0\left| {\bf E}%
_{\sigma }^{2}({\bf x}_{A},t)\right| 0\right\rangle .  \label{WA}
\end{equation}
In the last equation $\sigma =({\bf k},\lambda )$ characterizes the
electromagnetic field mode (wave vector and polarization state respectively)
and $\alpha _{A}(\omega _{\sigma })$ is the electric polarizability of atom $%
A$ at frequency $\omega _{\sigma }$. It is now convenient to express each
mode of the total electromagnetic field operator at the position of atom $A$
as the sum of two contributions: 
\begin{equation}
{\bf E}_{\sigma }({\bf x}_{A},t)={\bf E}_{0\sigma }({\bf x}_{A},t)+{\bf E}%
_{B\sigma }({\bf x}_{A},t),  \label{Etotal}
\end{equation}
where the first term on the r.h.s. of Eq. (\ref{Etotal}) stands for the
unconstrained vacuum field contribution and the second term represents the
contribution due to the presence of atom $B$. Since we are interested in the
interaction potential between the two atoms, we insert Eq. (\ref{Etotal})
into Eq. (\ref{WA}) and retain only the crossed terms. The dispersive van
der Waals interaction potential is then identified as 
\begin{equation}
U(r):=-{\frac{1}{2}}\sum_{\sigma }\alpha _{A}(\omega _{\sigma })\left\langle
0\left| {\bf E}_{0\sigma }^{{\bf (+)}}({\bf x}_{A},t)\cdot {\bf E}_{B\sigma
}({\bf x}_{A},t)+{\bf E}_{B\sigma }({\bf x}_{A},t)\cdot {\bf E}_{0\sigma }^{%
{\bf (-)}}({\bf x}_{A},t)\right| 0\right\rangle ,  \label{U}
\end{equation}
where we have decomposed the vacuum field into positive and negative
frequencies with 
\begin{equation}
{\bf E}_{0\sigma }^{{\bf (\pm )}}({\bf x},t)=\pm \;\,i\left( {\frac{2\pi
\hbar \omega _{\sigma }}{V}}\right) ^{1/2}a_{\sigma }^{(\mp )}(0)\exp (\mp
i\omega _{\sigma }t)\exp (\pm i{\bf k}\cdot {\bf x}){\bf \hat{e}}_{{\bf %
\sigma }},  \label{vacuumEfield}
\end{equation}
where $a_{\sigma }^{(-)}(0)$ is the anihilation operator and $a_{\sigma
}^{(+)}(0),$ the creator one; ${\bf \hat{e}}_{\sigma }$ is the polarization
vector. In order to obtain an approximate expression for the field operator $%
{\bf E}_{B\sigma }({\bf x}_{A},t)$ we first recall that the classical
electrical field at ${\bf x}_{A}$ generated by an oscillating magnetic
dipole ${\bf m}$ located at ${\bf x}_{B}$ is given by \cite{Jackson}: 
\begin{equation}
{\bf E}_{m}({\bf x}_{A},t)=-\left\{ {\frac{{\bf \dot{m}}(t-r/c)}{cr^{2}}}+{%
\frac{{\bf \ddot{m}}(t-r/c)}{c^{2}r}}\right\} \times {\bf \hat{r}},
\label{Em}
\end{equation}
where $r=|{\bf r}|:=|{\bf x}_{A}-{\bf x}_{B}|$. A comment is in order here:
notice that the electric field of an oscillating magnetic dipole does not
contain the static term, {\it i. e.}, there is no term proportional to $%
1/r^{3}$ in Eq. (\ref{Em}). Observe also that the equations for the
electromagnetic field operators in the Heisenberg picture are formally
identical to their classical counterparts. Hence, we can obtain a good
approximation for the operator field ${\bf E}_{B,\sigma }({\bf x}_{A},t)$ if
we think of ${\bf m}$ in Eq. (\ref{Em}) as the magnetic dipole of atom $B$
induced by the vacuum magnetic field, that is, if we write 
\begin{equation}
{\bf m}(t)=\sum_{\sigma }\beta _{B}(\omega _{\sigma })\left[ {\bf B}%
_{0\sigma }^{{\bf (+)}}({\bf x}_{B},t)+{\bf B}_{0\sigma }^{{\bf (-)}}({\bf x}%
_{B},t)\right] ,  \label{m}
\end{equation}
where $\beta _{B}(\omega _{\sigma })$ is the magnetic polarizability of the
atom $B$ at the frequency $\omega _{\sigma }$. Since the atom $A$ is only
electrically polarizable and atom $B$ is only magnetically polarizable, we
shall suppress from now on the subscripts $A$ and $B$ from $\alpha $ and $%
\beta $ respectively. The analogues of Eq. (\ref{vacuumEfield}) for the
vacuum magnetic field operators are given by: 
\begin{equation}
{\bf B}_{0\sigma }^{(\pm )}({\bf x},t)=\pm \;\,{i}\left( \frac{2\pi \hbar
c^{2}}{V\omega _{\sigma }}\right) ^{1/2}\;a_{\sigma }^{(\mp )}(0)\exp (\mp
i\omega _{\sigma }t)\;\exp (\pm i{\bf k}\cdot {\bf r}){\bf \hat{k}}\times 
{\bf \hat{e}}_{\sigma }  \label{vacuumBfield}
\end{equation}
Inserting Eqs. (\ref{vacuumBfield}) into (\ref{m}) and the result into Eq. (%
\ref{Em}) we get the electric field operator ${\bf E}_{B\sigma }({\bf x}%
_{A},t)$. Substituting this expression into Eq. (\ref{U}) we obtain the
following expression for $U(r)$: 
\begin{eqnarray}
U(r) &=&{\frac{1}{8\,\pi \,V}}\;{\cal R}e\left[ \sum_{\sigma }k^{3}\,\alpha
(\omega _{\sigma })\,\beta (\omega _{\sigma })\;\exp (-ikr)\exp \left[ i{\bf %
k}\cdot ({\bf x}_{A}-{\bf x}_{B})\right] \;\right.   \nonumber \\
&&\left. \times \left( {\frac{i}{(k\,r)^{2}}}-{\frac{1}{k\,r}}\right) \left(
\;{\bf \hat{e}}_{{\bf \sigma }}\cdot ({\bf k}\times {\bf \hat{e}}_{{\bf %
\sigma }})\times {\bf \hat{r}}\right) \right] .
\end{eqnarray}
Passing to the continuum $\sum_{{\bf k}\lambda }\rightarrow \;{\frac{V}{%
(2\pi )^{3}}}\;\sum_{\lambda =1}^{2}\int_{0}^{\infty }\;dk\;k^{2}\oint
\,d\Omega _{{\bf k}}$ we compute the angular integral and after a lengthy
calculation we obtain: 
\begin{equation}
U(r)={\frac{\hbar }{\pi \,c^{6}}}\;\int_{0}^{\infty }\,d\omega \,\omega
^{6}\,\alpha (\omega )\,\beta (\omega )\;{\cal G}\left( {\frac{\omega \,r}{c}%
}\right) \;,  \label{V2}
\end{equation}
where we have defined 
\begin{equation}
{\cal G}(x):=-{\frac{\sin (2x)}{x^{4}}}+2\,{\frac{\cos (2x)}{x^{3}}}+{\frac{%
\sin (2x)}{x^{2}}}\;.  \label{G}
\end{equation}
The last two equations give the general expression for the interaction
potential between an electrically polarizable atom and a magnetically
polarizable one. However, it is convenient to analyze the retarded and the
non-retarded limits separately. For large distances compared with the atomic
transition wavelengths ($r>>c/\omega _{mn}$), as in the case of two
electrically polarizable atoms, only the vacuum field modes with large
wavelengths are effective in polarizing the atoms and hence, as a first
approximation, we may replace the polarizabilities $\alpha (\omega )$ and $%
\beta (\omega )$ by their static values. Consequently, the retarded
interaction potential between these atoms is given by: 
\begin{equation}
U(r)={\frac{\hbar \,c\,\alpha (0)\beta (0)}{\pi r^{7}}}\int_{0}^{\infty
}\,dx\;x^{6}\;{\cal G}(x)\;.  \label{VR}
\end{equation}
Substituting Eq.(\ref{G}) into Eq.(\ref{VR}) and evaluating the needed
integrals, we obtain: 
\begin{equation}
U_{R}(r)={\frac{7\,\hbar \,c\,\alpha (0)\beta (0)}{4\,\pi \,r^{7}}}\;,
\label{VRfinal}
\end{equation}
which agrees with Eq.(\ref{UBoyer}) if in this equation we set $\alpha
_{2}=\beta _{1}=0$. Therefore, the force in the retarded case behaves like $%
1/r^{8}$, in perfect analogy with the Casimir and Polder result for two
electrically polarizable atoms.

For short distances ($r<<c/\omega _{mn}$), a situation where the retardation
effects can be neglected, one would naively expect that $U_{NR}(r)\propto
1/r^{6}$ (and hence $F_{NR}\propto 1/r^{7}$), but as we will show, this is
not so. In this limit, the dominant contribution to the integral in Eq.(\ref
{V2}) comes from the first term of the r.h.s. of Eq.(\ref{G}), so that: 
\begin{equation}
U_{NR}(r)={\frac{\hbar }{\pi \,c^{6}}}\int_{0}^{\infty }\,d\omega \,\omega
^{6}\,\alpha (\omega )\,\beta (\omega )\left\{ -{\frac{\sin \left( 2\omega
r/c\right) }{(\omega r/c)^{4}}}\right\} \;.  \label{VNR}
\end{equation}
In order to evaluate this integral, we need the expressions for $\alpha
(\omega )$ and $\beta (\omega )$. From perturbative quantum mechanics, it
can be shown that \cite{Davydov65} 
\begin{equation}
\alpha _{n}(\omega )={\frac{2}{3\,\hbar }}\,\sum_{m}\,{\frac{\omega _{mn}\,|%
{\bf d}_{mn}|^{2}}{\omega _{mn}^{2}-\omega ^{2}},}  \label{alpha}
\end{equation}
where ${\bf d}_{mn}$ is the transition (electric) dipole matrix element. An
analogous expression also holds for $\beta (\omega )$, if we replace ${\bf d}%
_{mn}$ by ${\bf \mu }_{mn}$ in Eq. (\ref{alpha}), ${\bf \mu }_{mn}$ being
the transition (magnetic) dipole matrix element. In Eq. (\ref{alpha}) we
have neglected the linewidths, but they do exist so that for $\omega $ equal
to the ressonance frequencies ($\omega =\omega _{mn}$) the real part of $%
\alpha $ vanishes (the same remarks hold for $\beta $), and the integral in
Eq. (\ref{VNR}) is indeed well behaved. With this in mind, we can write:

\begin{eqnarray}
U(r) &=&-{\frac{\hbar }{\pi \,c^{2}\,r^{4}}}\;{\cal I}m\;\int_{0}^{\infty
}\,d\omega \,\omega ^{2}\;\alpha (\omega )\,\beta (\omega )\;\exp (i2\omega
r/c)  \nonumber \\
&=&-{\frac{\hbar }{\pi \,c^{2}\,r^{4}}}\;{\cal I}m\;\int_{0}^{\infty
}\,i\,d\sigma \,(i\sigma )^{2}\;\alpha (i\sigma )\,\beta (i\sigma )\exp
(-2\omega r/c)  \nonumber \\
&=&+{\frac{\hbar }{\pi \,c^{2}\,r^{4}}}\;\left( {\frac{2}{3\hbar }}\right)
^{2}\,\sum_{m,p}\,\omega _{mn}\,\omega _{pn}\,|{\bf d}_{mn}|^{2}\,|{\bf \mu }%
_{mn}|^{2}\;\int_{0}^{\infty }\,{\frac{\sigma ^{2}\;\exp (-2\sigma r/c)}{%
\left( \sigma ^{2}+\omega _{mn}^{2}\right) \left( \sigma ^{2}+\omega
_{pn}^{2}\right) }}\;d\sigma ,
\end{eqnarray}
where we used the Cauchy residue theorem and also used Eq. (\ref{alpha}) as
well as the analogous equation for $\beta $. Since we are investigating the
short distance behavior of $U(r)$, it is legitimate to make the
approximation $\exp (-2\omega r/c)\approx 1$ in the previous integral
(though $\sigma $ is integrated from $0$ to $\infty $, the integ:rand
vanishes for large values of $\sigma $ due to the powers of $\sigma $
present in the denominator). Moreover, assuming for simplicity that there is
a dominant transition in each atom, the above equation takes the form: 
\begin{eqnarray}
V(r) &=&{\frac{\hbar }{\pi \,c^{2}}}\,\left( {\frac{2}{3\hbar }}\right)
^{2}\,{\frac{|{\bf d}|^{2}\,\left| {\bf \mu }\right| ^{2}}{r^{4}}}%
\;\int_{0}^{\infty }{\frac{\sigma ^{2}\;d\sigma }{\left( \sigma ^{2}+\omega
_{0\alpha }^{2}\right) \left( \sigma ^{2}+\omega _{0\beta }^{2}\right) }} 
\nonumber \\
&=&\frac{\hbar }{2c^{2}r^{4}}\frac{\omega _{0\alpha }^{2}\,\omega _{0\beta
}^{2}\,\alpha \beta }{\left( \omega _{0\alpha }+\omega _{0\beta }\right) },
\label{FinalResult}
\end{eqnarray}
where $\omega _{0\alpha }$ and $\omega _{0\beta }$ are the dominant electric
and magnetic transition frequencies, respectively, and $\alpha :=\omega
_{0\alpha }^{-1}(2/3\hbar )|{\bf d}|^{2}$ and $\beta :=\omega _{0\beta
}^{-1}(2/3\hbar )|{\bf \mu }|^{2}$ are the electric and the magnetic
polarizabilities. This result shows that when we go from the retarded to the
non-retarded regime there is a striking change from $r^{-7}$ ($r^{-8}$) to $%
r^{-4}$($r^{-5}$) in the dispersive interaction potential (force). This is
to be compared with the case where the two atoms are electrically
polarizable only, in which the change is from $r^{-7}$ ($r^{-8}$) to $r^{-6}$%
($r^{-7}$) in the potential (force). The reason for this dramatic change can
be traced back to the absence of the static term in the expression for the
electric field operator created by a magnetic dipole induced by the field
fluctuations. It is also the present authors ' opinion that such a change is
well worth an experimental investigation. Such a radical change could be
easier to measure than the purely electically polarizable case measured by
Tabor and Winterton \cite{Tabor69}. For example, it would be  interesting to
observe the transition from $r^{-8}$ to $r^{-5}$ in the dispersive force
between a hydrogen atom and a helium atom. The result obtained here can be
also of some relevance in the analysis of the force between two macroscopic
bodies.

\end{document}